\documentclass[letterpaper]{article} 
\usepackage{aaai2026} 
\nocopyright
\usepackage{times} 
\usepackage{helvet} 
\usepackage{courier} 
\usepackage[hyphens]{url} 
\usepackage{graphicx} 
\usepackage{natbib} 
\usepackage{caption} 
\usepackage{booktabs}
\usepackage{array}
\newcolumntype{L}[1]{>{\raggedright\arraybackslash}p{#1}}
\newcolumntype{C}[1]{>{\centering\arraybackslash}p{#1}}

\newcommand{\cog}{\textit{cognitive stewardship}}

\title{What Does the Credential Still Certify? Cognitive Stewardship for AI-Mediated Education}
\author{Kai Yao}
\affiliations{School of Informatics, University of Edinburgh\\
Edinburgh, United Kingdom\\
\texttt{kai.yao@ed.ac.uk}\\[0.75em]
{\small\textit{Accepted at the Ninth AAAI/ACM Conference on AI, Ethics, and Society (AIES 2026), Malm{\"o}, Sweden, October 12--14, 2026.}}}

\begin{document}
\maketitle

\begin{abstract}
Generative AI is changing a basic premise of educational assessment: that submitted work can reliably evidence the human capacities a credential claims to certify. The challenge is not simply whether students use AI, but what remains inferable about learning when some cognitive work has been delegated to a system. This paper develops \emph{cognitive stewardship}, a framework for AI-mediated assessment that links the learning claim, delegation boundary, evidence standard, and safeguards. We then audit verified public generative AI assessment guidance from 30 universities. Using a pre-specified scoring codebook--a written, source-grounded rubric--four open-weight LLM models applied the rubric as structured coders, with scores averaged to reduce dependence on any single model's bias. The audit shows that public policies are becoming better at classifying AI use than at explaining what evidence and protections preserve credential validity. Boundaries are more visible than evidence standards; safeguards are uneven; and guidance is clearest when AI use resembles final-output substitution rather than feedback, access, verification, or professional workflow. The takeaway is that permission categories are necessary but insufficient. Universities need policies that make the certification logic visible: what learners may delegate, what they must still demonstrate, and how institutions will protect fair evidence rather than merely monitor AI use.
\end{abstract}

\section{Introduction}

Generative AI has made a quiet premise of educational assessment newly fragile: that a submitted artifact can stand as evidence of a learner's competence. A polished essay, program, proof, lesson plan, literature review, or design proposal may still show understanding. It may also reflect intensive machine assistance, private coaching, hidden outsourcing, or a legitimate accessibility support that is difficult to reconstruct after submission. The resulting problem is not only misconduct. It is whether the work handed in still supports the human claim a grade, course, or credential makes.

This paper calls that problem \emph{educational delegation}. The key question is not whether AI touched the work, but which cognitive operations moved from the learner to the system and which remained with the learner. One student may use AI feedback while retaining problem formulation, source evaluation, revision judgment, and final responsibility. Another may delegate topic selection, evidence search, argument structure, drafting, citation, and prose revision. Both cases involve AI, but they support very different educational inferences. Assessment validity asks whether a task produces evidence for a learning claim; credential validity asks whether accumulated evidence justifies an institutional claim about the learner. Both depend on knowing what was delegated, what evidence remains, and what protections make that evidence fair to produce and judge \cite{Messick1989,Sadler1989,BoudFalchikov2006,Tai2018,Dawson2024,Corbin2025}.

The paper turns this validity problem into a policy question with four parts. A policy should state the \textit{learning claim} being certified, the \textit{delegation boundary} that tells learners what AI may do, the \textit{evidence standard} used to judge competence after delegation, and the \textit{safeguards} that protect the evidence process. Safeguards include privacy, accessibility, tool-access equity, workload proportionality, appeal or due process, caution about false accusations or AI detection, non-AI alternatives, and vendor governance. If the claim is source evaluation, for example, the boundary concerns whether AI may summarize or rank sources; the evidence might be an annotated source map or short defense; and the safeguard might be a privacy rule for uploaded sources or an accessible non-AI route.

The issue is already practical rather than speculative. Recent UK higher-education surveys motivate the governance problem without serving as global prevalence estimates: they report widespread student use of generative AI, substantial use for assessed work, uncertainty about acceptable use, and concern about false accusations, skill loss, access gaps, privacy, and employability \cite{HEPI2026,Jisc2025}. Universities are responding with AI-use categories, open and secure assessment lanes, task-level AI statements, disclosure requirements, and training \cite{UCLGenAI,SydneyAI,UQAI,AucklandAI,UNSWAI,BristolAI}. These responses show that institutions already recognize the limits of a universal ban or a blanket permission rule, even when they do not yet name the issue as delegation governance.

The learning evidence cuts in both directions. AI tutoring and feedback can support learning when they are embedded in active engagement, retrieval, formative guidance, feedback literacy, and self-explanation \cite{Bloom1984,HattieTimperley2007,NicolMacfarlaneDick2006,BoudMolloy2013,CarlessBoud2018,VanLehn2011,Ma2014,KulikFletcher2016,ChiWylie2014,Kestin2025,OECD2026Digital}. Earlier AI-in-education scholarship also cautions that technical systems must be interpreted through pedagogy, institutions, and educator practice rather than treated as autonomous solutions \cite{ZawackiRichter2019,RollWylie2016,WilliamsonEynon2020,Selwyn2024}. The same system can help one learner practice judgment and allow another to bypass the practice through which judgment develops.

Nor is offloading itself the enemy. Notebooks, calculators, search engines, and collaborators are ordinary parts of human cognition \cite{RiskoGilbert2016}. But offloading changes what is practiced and what can be inferred from performance. Automation research warns that assistance can produce complacency, miscalibrated trust, mode confusion, loss of situation awareness, and shallow memory for externally stored information \cite{Bainbridge1983,Endsley1995,SarterWoods1995,ParasuramanRiley1997,ParasuramanManzey2010,LeeSee2004,HoffBashir2015,Sparrow2011}. Recent work on generative AI and critical thinking similarly suggests that reliance can move human work from doing a task toward verifying and supervising outputs \cite{Lee2025}. That shift can be educationally valuable only when verification and supervision are themselves taught, evidenced, and assessed.

The argument should not rest on today's model failures. If AI systems become more accurate, safer, cheaper, and more capable, why should humans still learn? Many defenses of education rely on present weaknesses such as hallucination, bias, weak reasoning, limited context, or unreliability \cite{Bender2021,Kasneci2023,Lo2023,Dwivedi2023}. Those weaknesses matter, but they are an unstable foundation for long-term educational policy. The stronger question is which human capacities should remain visible even when task performance can be delegated.

The contribution is both conceptual and empirical. Conceptually, the paper defines educational delegation and develops \cog{} as a framework linking learning claims, delegation boundaries, evidence standards, and safeguards. This distinguishes credential-validity governance from learner-facing AI literacy, evaluative judgment, and AI-use scales. Empirically, the paper audits public university policies to examine whether public guidance connects permission rules to evidence and safeguards. Worked examples, tradeoff analysis, and scope conditions show how the framework can guide institutional revision without treating every AI use as either misconduct or progress.

\section{Educational Delegation}

Educational delegation names the relation among tool use, task design, and the certified claim. It asks what learners may assign to AI, what evidence still connects them to the claim, and what safeguards make that arrangement defensible. A policy that says only ``AI allowed'' or ``AI prohibited'' treats AI as a single kind of help. In practice, learners may ask AI to translate text, explain concepts, summarize sources, draft prose, debug code, generate examples, plan an argument, or evaluate a solution. Delegating spell-checking is not the same as delegating evidence selection. Translation for access is not the same as delegating the language skill being assessed.

Artifact-centered assessment is fragile for this reason. A final product can look excellent while revealing little about which operations the learner performed. Detection does not solve the problem. Even perfect detection of AI involvement would not show whether the use was educationally appropriate, whether it displaced the target capacity, or whether the learner could inspect the result. The institution needs a prior account of the learning claim; only then can it decide which delegation is acceptable, which evidence is needed, and which safeguard is proportionate.

The framework uses five recurring delegation types. \emph{Access support} enables participation, for example through translation, speech-to-text, or assistive scaffolding. \emph{Feedback support} offers critique or suggestions while the learner retains judgment. \emph{Process support} assists planning, drafting, debugging, or workflow. \emph{Substitution} occurs when AI performs the operation being assessed. \emph{Output-verification support} treats checking AI-mediated work as a capability to be taught and assessed, not merely as a safeguard after use. These categories describe the role of assistance in a task; they are not moral labels or fixed tool labels. The same use can be access support in one task, process support in another, and substitution in a third.

This framing complements assessment scales that classify permitted levels of AI use \cite{Furze2024,Perkins2025}. Those scales help communicate course rules. Cognitive stewardship asks the institutional question that follows: what claim is the rule protecting, and what evidence makes the claim defensible? Emerging university policies already move in this direction through category systems, secure and open assessment lanes, course-profile options, disclosure requirements, and task-level guidance \cite{UCLGenAI,SydneyAI,AucklandAI,UQAI,UNSWAI,BristolAI}. The supplement gives the full delegation taxonomy.

Credential validity is the central reason to take delegation seriously. Institutions certify people, not just products. That certification carries several educational commitments. Learners need enough understanding for agency and freedom from dependency \cite{Sen1999,Nussbaum2010}. Education should also prepare people to question rules, institutions, and automated decisions \cite{Dewey1916,Arendt1961,Biesta2010}. Delegated decisions can affect patients, clients, publics, and other learners \cite{Vallor2016,Santoni2018}. AI governance can also become surveillance, exclusion, or platform dependence \cite{Selbst2019,Mittelstadt2019,AnannyCrawford2018,Raji2020}. These commitments can conflict. A ban may protect foundational practice but harm access. Permission may support authentic workflow but weaken evidence of learner understanding. Delegation governance makes the tradeoff visible rather than hiding it behind a general attitude toward AI.

\section{Cognitive Stewardship Framework}

\textit{Cognitive stewardship} is the institutional practice of governing how cognitive work is divided between humans and AI while preserving understanding, verification, responsibility, access, and contestability. It builds on work about AI literacy, self-regulated learning, evaluative judgment, and human control in human-AI systems \cite{LongMagerko2020,Ng2021,Tai2018,Sperber2010,Amershi2019,Zhang2026}. Its unit of analysis is not an individual competency checklist, but the warrant behind a course, program, or credential. It asks when an institution may still certify a claim after AI-mediated production.

The framework aligns four elements: the \textbf{learning claim} being certified; the \textbf{delegation boundary} specifying what AI may do; the \textbf{evidence standard} showing what observable work, explanation, verification, or defense remains; and the \textbf{safeguards} protecting privacy, accessibility, proportionality, equity, and appeal. A policy is under-specified when any element is missing. A course can clearly permit AI and still fail to say what it can certify; it can prohibit AI and still fail to protect access or due process.

\begin{table*}[t]
\centering
\scriptsize
\setlength{\tabcolsep}{2.2pt}
\renewcommand{\arraystretch}{1.02}
\begin{tabular*}{\textwidth}{@{\extracolsep{\fill}}L{0.19\textwidth}L{0.23\textwidth}L{0.27\textwidth}L{0.25\textwidth}@{}}
\toprule
\textbf{Learning claim} & \textbf{Boundary} & \textbf{Evidence} & \textbf{Safeguard pattern} \\
\midrule
Foundational fluency & AI-free or tightly limited for representative sub-tasks. & In-class work, short explanations, error correction, and transfer to near cases. & Low stakes, accommodations, and clear fading criteria. \\
Feedback uptake & AI may critique, suggest, or pose alternatives while the learner retains judgment. & Revision memo, accepted and rejected suggestions, and peer or instructor discussion. & No private-data disclosure requirement; non-AI feedback option. \\
Professional workflow & AI may support authentic disciplinary practice. & Portfolio, provenance note, tests, defense of choices, and error review. & Tool access, privacy review, and disciplinary standards. \\
Output verification & Output-verification support is required rather than treated as optional policing. & Comparison across outputs, source verification, bias or error analysis, and escalation decisions. & Avoid punitive framing; reward uncertainty and correction. \\
High-stakes credential & Delegation depends on public risk and the specific capacity being certified. & Secure demonstration plus authentic AI-mediated task where appropriate. & Appeals, accessible alternatives, and proportional monitoring. \\
\bottomrule
\end{tabular*}
\caption{Illustrative delegation-centered assessment patterns. Boundary, evidence, and safeguard choices change with the certified claim: foundational fluency may require limited delegation, professional workflow may permit broad AI use with defense of choices, and output verification may itself become the assessed capability.}
\label{tab:model}
\end{table*}

Table~\ref{tab:model} illustrates the main design rule: start with the certified claim, not the tool. If delegation would remove the operation being certified, the boundary should be restrictive or the task should be redesigned. If delegation supports the claim but changes the evidence, the policy should add process evidence, explanation, verification, or defense. If delegation leaves the claim largely untouched, the policy should avoid unnecessary friction. These are validity judgments, not moral categories.

The limiting case clarifies the rule. Imagine a safe, accurate, cheap, and broadly capable system that can complete schoolwork, professional tasks, creative projects, and everyday planning. Some outsourcing would be rational: learners may seek efficiency, accommodation, rest, credentials, or employability. The institutional limit is narrower but firm. A credential cannot certify a human capacity while allowing unrestricted delegation of that very capacity. A student may use a calculator in advanced mathematics, but the institution still needs evidence of mathematical understanding. A programmer may use code generation, but the institution still needs evidence that the programmer can read, test, debug, and take responsibility for code.

Cognitive stewardship therefore treats validity as a warrant structure. The institution makes a claim about the learner; the assessment supplies evidence; the delegation boundary explains why AI assistance has not broken the link; safeguards explain why the evidence was produced and judged under fair conditions. When those pieces are aligned, AI can be part of valid assessment. When they are separated, policy may look clear while the credential inference remains weak.

\section{Examples and Governance Conditions}

Two examples show how the framework changes assessment design. In a first-year writing course, a traditional final essay may no longer support all the claims attached to it: source comprehension, argument formation, prose control, revision, and judgment about feedback. A stewardship redesign distributes evidence across diagnostic work, a source map, an AI-feedback workshop, the final essay, and a revision memo or conference. The boundary is strict where the course needs unaided diagnosis, permissive where feedback can support learning, and evidence-oriented where students must accept, reject, or explain AI suggestions. If the evidence is inconsistent--for example, a polished essay paired with weak source understanding--the first response should be an evidentiary check, conference, or revision request, not an automatic accusation. The supplement gives the fuller writing-course design and tradeoff tables.

An introductory programming course has a different boundary. A blanket ban on code generation may preserve syntax practice but ignore the professional reality of AI-mediated software work. Blanket permission creates the opposite problem: novices may submit code they cannot read, debug, test, or secure. A stewardship design can split the claims. Short AI-limited tracing and debugging tasks preserve mental models, while projects allow AI support for boilerplate, refactoring, documentation, and test generation if students explain design choices, run tests, identify AI-generated errors, and defend tradeoffs. The point is not to find one rule for all programming. It is to keep the evidence matched to the claim.

These examples also show why cognitive stewardship is not merely an assessment technique. It is a governance arrangement. Institutions procure tools, define legitimate help, collect disclosures, set penalties, and certify competence. Infrastructure becomes most consequential when it fades into everyday practice \cite{StarRuhleder1996}. Vendor lock-in can make schools dependent on platforms they cannot audit, and surveillance harms may fall unevenly on racialized, disabled, low-income, international, and linguistically marginalized learners when automated accusations or behavioral monitoring are layered onto existing suspicion \cite{Eubanks2018,Noble2018,Benjamin2019,CouldryMejias2019}. If students cannot access safe tools, legitimate delegation becomes unequal; if platform terms appropriate student writing or research data, privacy becomes an educational concern; if monitoring is the main safeguard, integrity may be purchased at the cost of trust.

The framework should be applied in a risk-tiered way. Here, low-stakes work means formative activity that does not by itself determine progression or certification; higher-stakes work carries substantial weight in those decisions. Everyday assignments still need adequate evidence when they materially support a learning claim, but the burden should remain proportionate. Low-stakes work may need only clear AI-use statements, source maps, short reflection memos, or in-class checkpoints. Higher-stakes credentials may require secure demonstrations, sampled oral defenses, provenance notes, accessibility review, and independent appeals. Programs can provide shared tools, templates, rubrics, appeal routes, staff training, and student partnership rather than shifting these burdens silently to instructors.

Accessibility is not an exception to stewardship; it is one of its conditions. AI or related tools may legitimately support screen reading, translation, speech-to-text, executive-function scaffolding, dyslexia support, anxiety scaffolding, or multilingual access. The relevant question is whether assistance bypasses the learning claim or enables access to it. A ban that preserves unaided fluency for some students while excluding disabled or multilingual learners is not valid stewardship. A fair policy states the claim in functional terms--source evaluation, argument structure, debugging, clinical judgment--rather than treating unaided production as automatically neutral and assisted production as automatically suspect.

\section{Empirical Audit of Policy Practice}

The framework leads to a descriptive empirical question: when universities publish guidance about generative AI in assessment, do they merely classify AI use, or do they also explain why the remaining work can still support a learning or credential claim? We examined that question through a policy audit of public generative AI assessment guidance from 30 universities. The audit treats public guidance as a reader-facing artifact: it asks what a student, instructor, or reviewer can see about the four elements of \cog{}--claim, boundary, evidence, and safeguard--from official public text.

\begin{figure*}[t]
\centering
\includegraphics[width=0.98\textwidth]{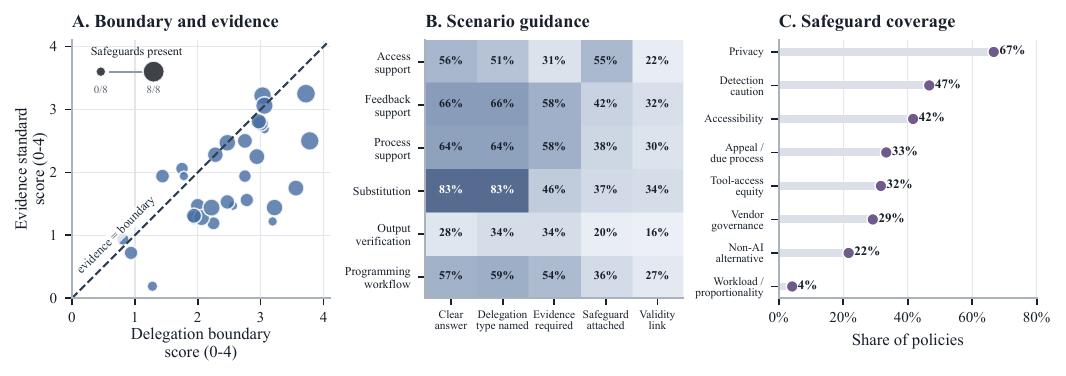}
\caption{Institutional AI guidance by boundary, evidence, scenario, and safeguard. Public guidance is stronger as permission policy than as credential-validity governance: boundary language exceeds evidence language, substitution is clearer than support-like uses, and safeguards appear unevenly.}
\label{fig:empiricalcore}
\end{figure*}

\noindent\textbf{Corpus, codebook, and scoring.}
The unit of analysis was a public institutional policy package: the set of official public sources through which one university tells students or instructors how generative AI may be used in assessed work. The purposive corpus covers five English-speaking national systems: the United Kingdom (11 packages), Australia (5), New Zealand (2), Canada (5), and the United States (7). Inclusion required a publicly reachable official institutional source, explicit relevance to generative AI in assessment or student work, and enough substantive guidance to apply the pre-specified rubric. Error pages, news-only items without operational guidance, and sources that could not be independently verified were excluded. The corpus was assembled to compare visible policy designs, not to estimate their prevalence worldwide. The supplement lists every official source entry point and the complete scoring protocol.

The audit used a pre-specified scoring codebook. Here, \emph{codebook} means a written rubric defining the policy constructs, score levels, scenario tests, and source-evidence rules used to translate policy text into comparable observations. A positive score required a quoted or locatable passage; unsupported higher scores were not counted. Four open-weight LLM models applied the same codebook to the same verified public text. Using four models reduced dependence on any one model's calibration, but the audit did not include an independently human-coded comparison set. Between-model agreement and standard deviation are therefore sensitivity measures, not validation: they cannot rule out errors shared across models or establish how expert human coders would score the text. We consequently treat exact score levels as exploratory descriptions and base interpretation on recurring patterns across constructs, scenarios, and source-grounded excerpts.

The institution-level constructs were learning claim, delegation boundary, evidence standard, and eight safeguards: privacy, accessibility, equity of tool access, workload/proportionality, appeal or due process, caution about false accusations or AI detection, non-AI alternatives, and vendor/procurement governance. The scale maxima differ because the constructs do different work. Learning claim is scored 0--3, boundary and evidence are scored 0--4, and safeguards are counted across eight indicators. Only boundary and evidence are directly compared as same-scale ordinal scores; safeguards are interpreted as coverage rather than severity.

For the scenario analysis, \emph{actionability} asks whether a student or instructor can tell what to do in a concrete use case. It is a 0--4 sum of four fields: clear answer, recognition of the delegation type, required evidence, and attached safeguard. A separate \emph{validity link} field records whether the policy explains why the answer preserves the learning or credential claim. Completeness combines claim, boundary, evidence, and safeguard coverage only to group visible policy profiles; because these components have different scales, it is not an overall quality score or an inferential ranking. The supplement reports model-level means, between-model standard deviations, and the full scoring instrument.

The analysis is intentionally conservative and public-text bounded. Because the codebook operationalizes the proposed framework, the audit is a diagnostic application of that normative lens rather than a framework-neutral test of institutional quality. A low score means that an element was not sufficiently visible in public text under the rubric; it does not establish weak internal practice. Figure~\ref{fig:empiricalcore} and Table~\ref{tab:empiricalpatterns} should therefore be read as exploratory diagnostics, not league tables.

\noindent\textbf{Boundary-evidence asymmetry.}
The first result is asymmetry rather than silence. Most audited packages drew some line around AI use: 24 reached a delegation-boundary score of at least 2, and the mean boundary score was 2.47/4. The weaker element was the evidence standard. Its mean was 1.89/4; 22 policies scored higher on boundary than evidence, 3 tied, and only 5 reversed the gap. Figure~\ref{fig:empiricalcore}A makes this imbalance visible because most points sit below the parity diagonal. The lower-right region is especially important: those packages classify AI assistance relatively clearly while giving much thinner public guidance about the evidence that should remain after assistance.

This asymmetry is the empirical form of the paper's central problem. Permission categories tell a reader whether an act is allowed, conditional, or prohibited. They do not by themselves warrant the inference from the submitted artifact to the learner's competence. A policy can say that AI may help with brainstorming, editing, or coding support; the credential question remains what the learner must still show. The audit finds that many public policies are better at classification than at warrant: they name the boundary more often than the evidence that would make the boundary educationally meaningful.

\noindent\textbf{Scenario guidance.}
The scenario stress test examines how each policy package applies to access support, feedback support, process support, substitution, output-verification support, and a domain-specific programming-workflow case. The first five scenarios correspond to the delegation types introduced earlier. Programming workflow is included because assessed coding work often combines code completion, debugging, explanation, test generation, and verification in one continuous task. For each policy-by-scenario unit, the scoring rule records whether the policy gives a clear answer, recognizes the delegation type, requires evidence, attaches a safeguard, and links the answer to assessment validity.

The results show why misconduct-centered policy is too narrow for delegation governance. Figure~\ref{fig:empiricalcore}B moves from basic clarity to educational justification. Substitution was the most actionable case, averaging 2.49/4 with clear answers in 83\% of packages, because final-output replacement fits familiar academic-integrity categories. Legitimate support cases were often conditionally covered, but evidence, safeguards, and validity links remained thinner. Policies may tell a reader whether a use is allowed while saying less about what trace of reasoning, revision, verification, or professional judgment should remain. Public guidance is therefore clearest when AI use resembles cheating and less complete when AI use resembles learning support, access support, or authentic professional workflow.

\begin{figure}[!t]
\centering
\includegraphics[width=0.92\columnwidth]{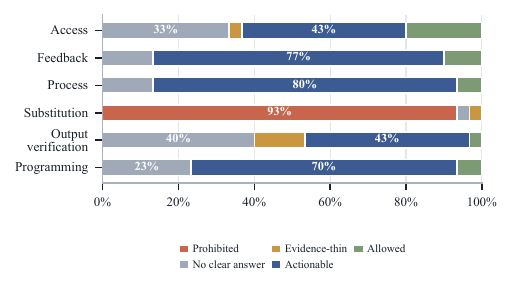}
\caption{Policy-answer categories across the six scenario stress tests. Evidence-thin conditions give a conditional answer without a usable evidence requirement; actionable conditions pair the condition with clearer evidence or safeguards.}
\label{fig:scenariooutcomes}
\end{figure}

\noindent\textbf{Outcome composition and safeguards.}
Figure~\ref{fig:scenariooutcomes} separates outcome labels from evidentiary strength. The key contrast is not allowed versus prohibited, but conditional permission with or without usable evidence. A policy can permit AI and remain weak if it gives no evidence standard. It can also be conditional and strong if it explains what the learner must still show. Feedback, process, output-verification, and programming-workflow cases often fall into this conditional space. Many institutions are therefore asking for context rather than simply allowing or banning these uses, but the required context is often not specified in terms of evidence. Conversely, the frequent prohibition of substitution does little for the more common cases in which AI supports brainstorming, revision, translation, code workflow, or error checking. Outcome labels are a starting point, not a validity argument.

The safeguard pattern in Figure~\ref{fig:empiricalcore}C explains why the evidence gap has practical consequences. Privacy was the most visible safeguard, appearing in about two thirds of packages. Detection caution and accessibility appeared in fewer than half; appeal or due process, tool-access equity, and vendor governance appeared in roughly one third; non-AI alternatives were rarer; workload/proportionality was almost absent. Safeguards are therefore not yet developing as a visible bundle. Public guidance is more likely to warn about privacy than to provide recourse, accessible alternatives, or proportional workload limits. The design risk is that disclosure and monitoring can make students more visible to the institution without making assessment fairer.

\begin{table*}[t]
\centering
\footnotesize
\begingroup
\renewcommand{\arraystretch}{1.08}
\begin{tabular*}{\textwidth}{@{\extracolsep{\fill}}L{0.245\textwidth}L{0.225\textwidth}L{0.245\textwidth}L{0.205\textwidth}@{}}
\toprule
\textbf{Analytic claim} & \textbf{Audit measure} & \textbf{Observed pattern} & \textbf{Interpretation} \\
\midrule
\textbf{Permission rules are more developed than evidence rules.} & Delegation-boundary and evidence-standard scores. & Mean boundary score was 2.47/4; mean evidence score was 1.89/4. 22 policies scored higher on boundary than evidence, 3 tied, and 5 reversed the gap. & AI-use categories are more visible than the evidence standards that would justify grades or credentials. \\
\midrule
\textbf{Safeguards are present but sparse.} & Count of privacy, accessibility, equity, workload, appeal, detection caution, non-AI alternative, and vendor-governance safeguards. & Packages contained a mean of 2.75 of 8 possible safeguards. The most visible categories were privacy at 67\%; detection caution at 47\%; accessibility at 42\%; appeal or due process at 33\%; equity of tool access at 32\%; vendor governance at 29\%; non-AI alternatives at 22\%; workload/proportionality at 4\%. & Public guidance often asks for disclosure or compliance without equally visible protection, recourse, or alternatives. \\
\midrule
\textbf{Policies are clearest for final-output substitution.} & Scenario actionability, a 0--4 score combining clear answer, delegation-type recognition, evidence requirement, and safeguard attachment. & Substitution averaged 2.49/4; the other five scenarios ranged from 1.17 to 2.32. & Policies handle obvious final-output replacement better than legitimate support and workflow delegation. \\
\midrule
\textbf{Policy clarity varies by missing design element.} & Completeness and internal-alignment scores. & 9 packages were higher-clarity; 4 were boundary-forward; 8 had limited evidence visibility; 9 were partially connected. & The framework identifies a design gradient rather than a binary compliant/noncompliant status. \\
\bottomrule
\end{tabular*}
\endgroup
\caption{Empirical patterns in the 30-institution policy audit. Boundary language exceeds evidence language, safeguards are sparse, substitution is the clearest scenario, and packages vary by missing design element rather than by a simple permissive/restrictive divide.}
\label{tab:empiricalpatterns}
\end{table*}

Table~\ref{tab:scenariosummary} reports the same scenario stress test numerically. It distinguishes three questions that policy discussion often collapses. \emph{Direct coverage} asks whether the scenario is named or strongly implied. \emph{Actionability} asks whether a reader can tell what to do, what type of delegation is involved, what evidence is required, and what safeguard applies. Higher actionability means easier application, not greater permissiveness. \emph{Validity link} asks whether the policy explains why that answer preserves the learning or credential claim. Across scenarios, public guidance is often able to classify an AI use, but much less often explains what makes the resulting assessment evidence valid.

\begin{table*}[t]
\centering
\scriptsize
\setlength{\tabcolsep}{1.6pt}
\begin{tabular*}{\textwidth}{@{\extracolsep{\fill}}L{0.135\textwidth}C{0.085\textwidth}C{0.080\textwidth}C{0.075\textwidth}C{0.070\textwidth}C{0.075\textwidth}C{0.075\textwidth}L{0.225\textwidth}@{}}
\toprule
\textbf{Scenario} & \textbf{Actionability (0--4)} & \textbf{Direct coverage} & \textbf{Clear answer} & \textbf{Evidence} & \textbf{Safeguard} & \textbf{Validity link} & \textbf{Most common outcome} \\
\midrule
Access support & 1.93 & 63\% & 56\% & 31\% & 55\% & 22\% & Actionable conditions (43\%) \\
Feedback support & 2.32 & 80\% & 66\% & 58\% & 42\% & 32\% & Actionable conditions (77\%) \\
Process support & 2.25 & 80\% & 64\% & 58\% & 38\% & 30\% & Actionable conditions (80\%) \\
Substitution & 2.49 & 93\% & 83\% & 46\% & 37\% & 34\% & Prohibited (93\%) \\
Output verification & 1.17 & 23\% & 28\% & 34\% & 20\% & 16\% & Actionable conditions (43\%) \\
Programming workflow & 2.07 & 40\% & 57\% & 54\% & 36\% & 27\% & Actionable conditions (70\%) \\
\bottomrule
\end{tabular*}
\caption{Scenario actionability and validity measures across 180 policy-by-scenario observations. Actionability is the mean 0--4 score combining clear answer, delegation-type recognition, evidence requirement, and safeguard attachment; the validity link column is reported separately and is not part of that sum. The most common outcome column reports the plurality category, not necessarily a majority.}
\label{tab:scenariosummary}
\end{table*}

\noindent\textbf{Archetypes and policy profiles.}
The archetype analysis translates the scores into policy-design profiles. We assigned archetypes by asking which part of the four-part relation was most visibly missing or unstable: evidence, safeguards, or the connection among claim, boundary, evidence, and safeguards. Nine packages were higher-clarity cases, four were boundary-forward cases, eight had limited evidence visibility, and nine were partially connected cases. Table~\ref{tab:archetypes} summarizes the groups. Higher clarity did not mean more permissiveness. It meant that public text more often connected a permission or prohibition to evidence and safeguards. Boundary-forward packages show the main failure mode in a specific form: visible AI-use categories with thinner evidence and safeguard conditions.

\begin{table*}[t]
\centering
\small
\begin{tabular*}{\textwidth}{@{\extracolsep{\fill}}L{0.155\textwidth}C{0.060\textwidth}L{0.355\textwidth}L{0.345\textwidth}@{}}
\toprule
\textbf{Policy archetype} & \textbf{Count} & \textbf{Observed pattern} & \textbf{Design implication} \\
\midrule
Higher clarity & 9 & Boundaries are accompanied by comparatively more evidence or safeguard language. & AI guidance can be specific without reducing policy to prohibition. \\
Boundary-forward & 4 & Permitted and prohibited uses are clearer than the safeguards that legitimate the boundary. & Category systems need paired protection, recourse, accessibility, and proportionality language. \\
Limited public evidence visibility & 8 & Guidance exists, but the four-LLM scoring did not consistently connect it to evidence standards or safeguards. & An official AI page is not sufficient if learners and instructors cannot infer what evidence remains valid. \\
Partially connected & 9 & Some framework elements are visible, but no stable boundary-evidence-safeguard profile appears. & Institutions can improve by targeting the missing element rather than rewriting the entire policy regime. \\
\bottomrule
\end{tabular*}
\caption{Policy archetypes identified in the audit. The archetypes show that the empirical pattern is not a simple permissive-versus-restrictive divide; the key distinction is whether policy boundaries are connected to evidence and safeguards.}
\label{tab:archetypes}
\end{table*}

\begin{figure*}[t]
\centering
\includegraphics[width=0.95\textwidth]{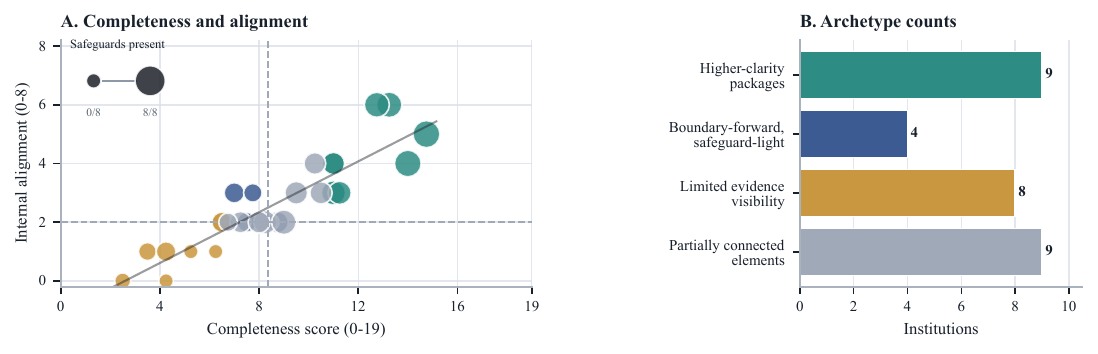}
\caption{Policy archetype scores and group counts for the 30-institution corpus. Packages are positioned by visible completeness and internal alignment; point size represents safeguard count. The figure shows policy clarity as a design gradient rather than a permissive/restrictive divide.}
\label{fig:archetypemap}
\end{figure*}

Figure~\ref{fig:archetypemap} relates the archetypes to the underlying scores; the supplement gives model-level variation, archetype diagnostics, and scenario score anchors. The map is descriptive, not a statistical model: it asks whether policy elements tend to appear together in public guidance. Movement from the lower-left toward the upper-right indicates that packages with more visible elements also tend to connect them more consistently. The lower-left cases may have official guidance but little public evidence-standard or safeguard language; the middle cases contain useful elements but unstable connections.

Institution-level cases make the archetypes concrete. To separate source transparency from score-bearing institutional comparison, Table~\ref{tab:institutioncases} reports six anonymized policy profiles. The named source index remains available in the supplement for corpus inspection, but the profile table focuses on contrasts in public policy design rather than institutional reputation.

\begin{table*}[t]
\centering
\scriptsize
\setlength{\tabcolsep}{2.2pt}
\begin{tabular*}{\textwidth}{@{\extracolsep{\fill}}L{0.145\textwidth}C{0.070\textwidth}C{0.078\textwidth}C{0.078\textwidth}C{0.070\textwidth}L{0.475\textwidth}@{}}
\toprule
\textbf{Policy profile} & \textbf{Learning claim (0--3)} & \textbf{Delegation boundary (0--4)} & \textbf{Evidence standard (0--4)} & \textbf{Safeguards (0--8)} & \textbf{Why this profile is shown} \\
\midrule
Broadly connected & 1.75 & 3.75 & 3.25 & 6.00 & Illustrates a package in which claim, boundary, evidence, and safeguard language are all visible. \\
Boundary-rich & 2.00 & 3.75 & 2.50 & 5.75 & Shows clear permitted, conditional, and prohibited uses, with less developed evidence language. \\
Evidence-rich & 2.00 & 3.00 & 3.25 & 5.00 & Shows comparatively strong public signals about what evidence remains after AI-mediated work. \\
Balanced visibility & 2.00 & 3.00 & 3.00 & 4.75 & Pairs AI-use guidance with relatively strong evidence language and several safeguards. \\
Boundary-evidence gap & 1.75 & 3.25 & 1.25 & 0.00 & Clear categories are not paired with comparably visible evidence or safeguard language. \\
Limited public visibility & 0.50 & 1.25 & 0.25 & 0.50 & Official guidance exists, but public support for evidence and safeguard constructs remains limited under the rubric. \\
\bottomrule
\end{tabular*}
\caption{Anonymized policy profiles selected to anchor six contrasts used in the archetype analysis. Scores are four-LLM means on the scales shown and describe public-text visibility under the rubric, not institutional practice or teaching quality.}
\label{tab:institutioncases}
\end{table*}

\noindent\textbf{Policy profiles and scenario-level gaps.}
The profiles clarify that the audit is not measuring how often a university permits AI. The more connected profiles make it easier to relate an AI-use category to a task, an evidence expectation, and surrounding conditions. Their strength is not maximum permissiveness, but a clearer account of what remains assessable after AI mediation. The boundary-evidence-gap profile illustrates the opposite risk: a category system can look administratively mature while leaving evidence standards and safeguards thin.

The scenario gaps show the same boundary-evidence problem at a finer grain. Feedback and process support were directly covered in 80\% of packages, but evidence requirements and safeguards were lower. Students may be told that brainstorming, outlining, editing, or feedback are conditionally acceptable while the document says less about what trace of reasoning, revision, or source judgment should remain. Access support has a different imbalance. Its safeguard rate was higher than its evidence rate, reflecting some visible attention to accommodation and inclusion, but only 31\% of packages required evidence that would connect the assistance back to the assessed claim. Output-verification support was the thinnest scenario. Only 23\% of packages directly covered it, even though the capacity to inspect, challenge, and correct AI output is one of the strongest reasons to teach with AI.

The programming-workflow case shows why domain-specific examples are needed. General categories such as drafting, editing, generating, or submitting do not map cleanly onto code completion, explanation, debugging, test generation, and error interpretation. A policy that only asks whether AI wrote the final answer misses the more important evidence question: can the learner explain the design choice, inspect generated code, interpret tests, and identify failures? The policy gap is therefore not only low coverage, but a mismatch between outcome categories and educational inference.

\noindent\textbf{Uncertainty in public wording.}
Disagreement among the four LLM-coder scores is used as a flag for scoring sensitivity under the protocol. Mean standard deviation was 0.57 for delegation-boundary scores, 0.53 for evidence-standard scores, and 0.42 for learning-claim scores. Scenario application was similarly uneven: high-variation cases appeared in 13 to 22 of the 30 policy packages depending on the scenario. Low means indicate little source-based support for the construct. High standard deviations indicate wording that supported different source-grounded readings. This does not prove that human readers would disagree in the same way, but it marks places where public wording may force policy users to do more interpretive work.

The audit measures public policy design, not classroom practice, institutional intention, or learning outcomes. Its purposive corpus covers public English-language guidance from five higher-education systems, so the results should be read as a structured pilot rather than a representative survey. Official-source verification establishes provenance; between-LLM variation only marks sensitivity to model choice and wording. Within that scope, institutional guidance already exhibits the gap the paper analyzes: public AI-use rules often do not specify how those rules preserve credential validity.

\noindent\textbf{Empirical takeaway.}
The empirical results are most useful as a revision guide. First, they separate the existence of AI guidance from the quality of the inference that guidance supports. A university can have a public AI page, a disclosure rule, and a misconduct warning while still leaving the reader unsure what the institution can infer from assessed work. This is why the boundary-evidence scatter, the scenario matrix, and the outcome distribution are complementary rather than redundant. The scatter shows an institution-level gap; the scenario matrix shows where the gap appears in concrete uses; the outcome distribution shows that the same policy can prohibit substitution while leaving support-like uses only conditionally governed.

Second, the audit identifies revision priorities. For a boundary-forward package, the next step is rarely another category label. It is an evidence standard: what would show that the learner still met the claim after AI assistance? For a safeguard-light package, the next step is protection and recourse: what happens if disclosure creates privacy risk, if detection is wrong, if access support is mistaken for misconduct, or if students cannot use the same paid tools? For a package with limited evidence visibility, the next step is a shared public vocabulary, because students and instructors cannot apply a rule that only appears as scattered advice. This is also why the completeness score is not reported as a ranking. The useful result is the missing element, not the ordinal position of an institution.

Third, credential validity cannot be repaired at the end of the pipeline. If evidence is absent, later enforcement cannot reconstruct learner contribution without suspicion, workload, or unequal burden. If safeguards are absent, disclosure may make students more legible without making assessment fairer. Institutions therefore need to decide before assessment which forms of delegation are part of the learning claim, which merely assist access or workflow, and which require additional evidence.

\section{Implications and Scope}

These findings make the framework useful for revision rather than ranking. Scores identify missing design elements, not institutional virtue. A boundary-forward policy needs evidence design; a safeguard-light policy needs clearer protection, recourse, accessibility, and proportionality; a package with limited public evidence visibility needs a shared vocabulary before it can fairly discipline students or support instructors. The practical question is whether a reader can infer what the institution still claims about learner competence.

\noindent\textbf{Evidence before enforcement.}
Evidence standards should be stated at the level of the claim. If the claim is source evaluation, evidence might include a source map or short defense; if the claim is programming judgment, it might include a code walkthrough, test interpretation, or explanation of AI-generated bugs. The point is not to make every assignment heavier. It is to stop treating a final artifact as self-explanatory proof when the path to that artifact has changed.

\noindent\textbf{Safeguards as validity conditions.}
Safeguards should be treated as part of validity rather than as compliance afterthoughts. Privacy, appeal, accessibility, non-AI alternatives, tool access, workload limits, and vendor review affect what learners can safely do and what instructors can fairly infer. A disclosure rule without privacy guidance can create risk without improving evidence. A detection rule without appeal can produce suspicion without trust. Public clarity matters for the same reason: students should not have to infer institutional values from scattered warnings, and instructors should not have to translate broad statements into high-stakes judgments alone.

The framework should remain proportional. Employment, efficiency, rest, accommodation, and credential access are legitimate educational concerns. Many uses of AI feedback, translation, brainstorming, code explanation, and practice generation will be appropriate when they support rather than replace the claim being certified. Productive struggle matters when it builds durable understanding \cite{BjorkBjork2011,Dunlosky2013,Kirschner2006}; unnecessary friction, exclusion, and surveillance undermine that aim.

\noindent\textbf{Program-level policy review.}
A common first response to generative AI is to ask each instructor to write a course policy. That is necessary but insufficient. If every course invents its own categories, students face a patchwork of local rules and instructors carry the whole burden of interpretation. A program-level review can instead begin from a small set of credential claims. What should a graduate be able to do without assistance? What should they be able to do with ordinary professional tools? What should they be able to inspect or challenge when AI has shaped the output? Once those claims are named, individual assignments can vary without becoming incoherent.

This matters most for borderline cases. Feedback support, translation, planning, debugging, and code assistance are not naturally located on a single permissive--restrictive line. They can be good evidence in one assessment and poor evidence in another. A translation tool may enable a multilingual student to show source interpretation, or it may bypass a language-learning claim. AI-generated tests may be a legitimate programming aid if the claim is design judgment, but weak evidence if the claim is writing tests. The framework does not answer these questions by listing tools. It asks what human judgment remains visible after the tool has been used.

The same logic applies to enforcement. If a policy only says that students must disclose AI use, disclosure becomes a broad compliance signal rather than evidence for a claim. A stronger policy explains when disclosure is enough, when process evidence is needed, when an oral check or secure demonstration is proportionate, and when the appropriate response is accommodation or redesign rather than discipline. This is the practical difference between monitoring AI use and governing educational delegation.

Operationally, stewardship can be implemented through a short public template rather than a long rulebook. For each high-stakes task, the template can ask four questions: what capacity is certified, what help is allowed, what evidence remains, and what protection or appeal route applies? The template should also say when no extra evidence is needed. That negative case matters. Without it, institutions may turn every ordinary AI interaction into a compliance event, increasing workload without improving validity.

This division of labor prevents fairness questions from becoming hidden instructor discretion: central units set procurement, privacy, accessibility, and appeal rules; programs define credential claims; instructors choose proportionate task evidence; and students know the route before submission.

Credential validity is therefore also a public AI-governance concern. When institutions certify learning, they allocate opportunity, status, trust, and responsibility. If AI-mediated work makes the relation between artifact and learner unclear, the institution cannot repair that relation by detection alone. It has to say what kind of human participation matters and how that participation can be evidenced without turning education into surveillance. Tools will change, but that certification question is more stable: if a model becomes better at drafting, feedback, or code generation, the institution still asks whether the learner can evaluate, adapt, contest, and take responsibility for the result.

\section{Conclusion}

Generative AI changes what educational institutions can validly certify when learners may delegate parts of the work. The answer is not simply prohibition, permission, detection, or outsourcing. This paper has framed the problem as educational delegation and proposed cognitive stewardship: connect the learning claim, delegation boundary, evidence standard, and safeguard layer before treating a product as evidence of competence.

The policy audit supports that diagnosis. The audited universities were not silent about generative AI; many had official pages, AI-use categories, and disclosure language. The gap was specific: rules about allowed use outpaced evidence for what credentials still certify. Boundary scores exceeded evidence scores for most policy packages, scenario guidance was clearest for final-output substitution, and safeguards appeared unevenly. In short, institutions are learning to classify AI use faster than they are learning to explain the evidence and protections that make classification educationally valid.

That finding also clarifies the paper's contribution. Cognitive stewardship is not another name for AI literacy, nor a request for more disclosure paperwork. It is an institutional design test: can the policy explain the relation among the capacity being certified, the AI-mediated work allowed or disallowed, the evidence that remains, and the protections that make the inference fair? Longer policies are not automatically better. Clearer policies let a reader trace the route from permitted help to remaining evidence, and from evidence gathering to fair process.

For students and instructors, the test of a policy is ordinary and practical. A student should know before submission which forms of help are legitimate, which forms change the evidence burden, and what process exists if access, privacy, or disability concerns make a standard rule inappropriate. An instructor should know what to ask for when AI assistance is disclosed and when the right response is redesign rather than discipline. Those expectations do not require a new rule for every model. They require a stable account of what the institution wants assessed work to show.

The broader significance is that AI policy should be treated as part of assessment design, not as a technology-use appendix. Once assistance is framed as delegation, the policy question becomes more precise. The problem is not to make every act of tool use visible to the institution. It is to decide which acts matter for a particular certification claim, which acts are ordinary support, and which acts require additional evidence or protection. That framing also makes the fairness problem clearer. Evidence requirements, disclosure duties, secure demonstrations, and appeals should be designed together rather than added after suspicion arises.

This also explains why the framework is meant to be proportional. A low-stakes brainstorming activity may need no more than clear permission. A capstone project may need a provenance note, testing record, or defense of design choices. A high-stakes professional credential may need a secure demonstration alongside an authentic AI-mediated task. The common thread is not more paperwork. It is a better match between the claim, the permitted help, and the evidence left behind. Without that match, disclosure and detection can expand institutional control while leaving the credential inference unchanged.

The audit is limited to a purposive, English-language corpus of public policy documents from five national systems. It does not show classroom practice, institutional intent, outcomes, or all settings, and its LLM-coded scores lack an independent human-coded validation set. Its value is diagnostic: it identifies visible patterns that future human-coded and course-level studies can test. That work should examine whether revision memos, source maps, code walkthroughs, oral explanations, and output-verification tasks improve learning, trust, dispute resolution, and access, and should include students as policy readers because rules clear to administrators may still be opaque or burdensome in ordinary study.

The immediate policy lesson is modest but demanding. Universities do not need to predict every new model or forbid every use that complicates grading. They need an explicit certification claim and evidence design showing what the learner did, what was legitimately delegated, and how the learner remained able to interpret, contest, or take responsibility for the result. That explicit link--claim, delegation, evidence, and safeguard--is the standard the framework asks policy to make visible. No shortcut replaces that evidence. As models change, the exact boundary will move; the need for a defensible warrant will not.

That warrant also avoids two false choices. Institutions need not treat all AI assistance as cheating, because some delegation supports access, feedback, and authentic professional practice. Nor should they treat all AI-mediated productivity as learning, because some delegation removes the very operation being certified. The durable task is to make those differences legible before assessment, so that students can use legitimate help without fear, instructors can ask for evidence without improvising suspicion, and credentials can continue to mean more than successful artifact production.

A practical revision can begin with one high-stakes assignment or one program outcome. Name the capacity, list the forms of AI help that do and do not threaten it, decide what evidence is proportionate, and specify the protection or appeal route. Repeating that exercise across a program would give students a coherent map rather than a patchwork of warnings. That is the governance work a credential now requires.

\clearpage
\raggedbottom
\section{Ethical Considerations, Positionality, and Adverse Impact}

This paper concerns institutional policy design rather than an intervention with human subjects. The empirical audit used public university documents, verified official sources, and source-grounded excerpts; it did not collect student records, instructor evaluations, or classroom behavior. Its ethical risk is not participant exposure but overclaiming from public policy text. To reduce that risk, the paper treats the audit as a descriptive policy study, distinguishes public policy design from actual practice, and reports source verification, scoring rationales, and uncertainty summaries in the supplement.

The framework is written from the standpoint of institutional governance and assessment validity, a standpoint that can understate how students experience AI rules as labor, suspicion, access, or exclusion. The argument treats disability access, multilingual participation, workload, appeal, privacy, and procurement as design conditions rather than secondary considerations. The paper does not assume that unaided performance is neutral, that all learners need the same form of agency, or that high-resource universities define the global baseline for responsible AI education.

The main adverse impact risk is that cognitive stewardship could be implemented as more surveillance, paperwork, or instructor workload. That would contradict the framework. Stewardship should use proportionate evidence, low-burden templates, accessible alternatives, student participation, and appeal routes, with high-cost checks reserved for high-stakes claims. A second risk is that institutions could cite the framework to preserve outdated tasks. The intended use is narrower: protect capacities that remain educationally and publicly consequential, redesign tasks that AI has made poor evidence, and revise rules when they create exclusion, false accusation, or dependency.

The policy audit used a documented four-LLM scoring procedure as part of the empirical analysis of policy texts. The open-weight LLM models were applied only to verified public policy text under a shared scoring protocol and scoring codebook--a written rubric for translating policy text into defined scores and scenario judgments. They were not treated as sources, authors, or authorities. Source verification, protocol design, aggregation choices, prose, and claims remain the responsibility of the author.

\begin{samepage}
\section*{LLM Usage Declaration}

During manuscript preparation, the author used large language models (LLMs) and AI agents as assistive tools for literature searching, drafting scripts for the audit experiments, generating plots to visualize the data, and refining and polishing the language and structure of the text. The author has manually reviewed all AI-assisted outputs and verified their correctness and consistency, including the text, code, tables, and figures. The author takes full responsibility for the content of this manuscript.

\end{samepage}

\section*{Acknowledgments}

This work was supported by the Edinburgh International Data Facility (EIDF) and the Data-Driven Innovation Programme at the University of Edinburgh. The author also gratefully acknowledges the School of Informatics at the University of Edinburgh for facilitating access to EIDF computational resources.


\begin{thebibliography}{99}

\bibitem[Amershi et~al.(2019)]{Amershi2019} Amershi, S.; Weld, D.; Vorvoreanu, M.; Fourney, A.; Nushi, B.; Collisson, P.; Suh, J.; Iqbal, S.; Bennett, P. N.; Inkpen, K.; Teevan, J.; Kikin-Gil, R.; and Horvitz, E. 2019. Guidelines for human-AI interaction. In \textit{Proceedings of the 2019 CHI Conference on Human Factors in Computing Systems}, 1--13. \url{https://doi.org/10.1145/3290605.3300233}.
\bibitem[Ananny and Crawford(2018)]{AnannyCrawford2018} Ananny, M.; and Crawford, K. 2018. Seeing without knowing: Limitations of the transparency ideal and its application to algorithmic accountability. \textit{New Media \& Society}, 20(3):973--989.
\bibitem[Arendt(1961)]{Arendt1961} Arendt, H. 1961. The crisis in education. In \textit{Between Past and Future: Six Exercises in Political Thought}, 173--196. New York: Viking Press.
\bibitem[Bainbridge(1983)]{Bainbridge1983} Bainbridge, L. 1983. Ironies of automation. \textit{Automatica}, 19(6):775--779.
\bibitem[Bender et~al.(2021)]{Bender2021} Bender, E. M.; Gebru, T.; McMillan-Major, A.; and Shmitchell, S. 2021. On the dangers of stochastic parrots: Can language models be too big? In \textit{Proceedings of the 2021 ACM Conference on Fairness, Accountability, and Transparency}, 610--623. \url{https://doi.org/10.1145/3442188.3445922}.
\bibitem[Benjamin(2019)]{Benjamin2019} Benjamin, R. 2019. \textit{Race after Technology: Abolitionist Tools for the New Jim Code}. Cambridge: Polity.
\bibitem[Biesta(2010)]{Biesta2010} Biesta, G. 2010. \textit{Good Education in an Age of Measurement: Ethics, Politics, Democracy}. Boulder, CO: Paradigm Publishers.
\bibitem[Bjork and Bjork(2011)]{BjorkBjork2011} Bjork, E. L.; and Bjork, R. A. 2011. Making difficulties desirable in learning. In \textit{Psychology and the Real World: Essays Illustrating Fundamental Contributions to Society}, 56--64. New York: Worth Publishers.
\bibitem[Bloom(1984)]{Bloom1984} Bloom, B. S. 1984. The 2 sigma problem: The search for methods of group instruction as effective as one-to-one tutoring. \textit{Educational Researcher}, 13(6):4--16.
\bibitem[Boud and Falchikov(2006)]{BoudFalchikov2006} Boud, D.; and Falchikov, N. 2006. Aligning assessment with long-term learning. \textit{Assessment \& Evaluation in Higher Education}, 31(4):399--413.
\bibitem[Boud and Molloy(2013)]{BoudMolloy2013} Boud, D.; and Molloy, E. 2013. Rethinking models of feedback for learning: The challenge of design. \textit{Assessment \& Evaluation in Higher Education}, 38(6):698--712.
\bibitem[Carless and Boud(2018)]{CarlessBoud2018} Carless, D.; and Boud, D. 2018. The development of student feedback literacy: Enabling uptake of feedback. \textit{Assessment \& Evaluation in Higher Education}, 43(8):1315--1325.
\bibitem[Chi and Wylie(2014)]{ChiWylie2014} Chi, M. T. H.; and Wylie, R. 2014. The ICAP framework: Linking cognitive engagement to active learning outcomes. \textit{Educational Psychologist}, 49(4):219--243.
\bibitem[Corbin et~al.(2025)]{Corbin2025} Corbin, T.; Bearman, M.; Boud, D.; and Dawson, P. 2025. The wicked problem of AI and assessment. \textit{Assessment \& Evaluation in Higher Education}, 1--17. \url{https://doi.org/10.1080/02602938.2025.2553340}.
\bibitem[Couldry and Mejias(2019)]{CouldryMejias2019} Couldry, N.; and Mejias, U. A. 2019. \textit{The Costs of Connection: How Data Is Colonizing Human Life and Appropriating It for Capitalism}. Stanford, CA: Stanford University Press.
\bibitem[Dawson et~al.(2024)]{Dawson2024} Dawson, P.; Bearman, M.; Dollinger, M.; and Boud, D. 2024. Validity matters more than cheating. \textit{Assessment \& Evaluation in Higher Education}, 49(7):1005--1016. \url{https://doi.org/10.1080/02602938.2024.2386662}.
\bibitem[Dewey(1916)]{Dewey1916} Dewey, J. 1916. \textit{Democracy and Education: An Introduction to the Philosophy of Education}. New York: Macmillan.
\bibitem[Dunlosky et~al.(2013)]{Dunlosky2013} Dunlosky, J.; Rawson, K. A.; Marsh, E. J.; Nathan, M. J.; and Willingham, D. T. 2013. Improving students' learning with effective learning techniques: Promising directions from cognitive and educational psychology. \textit{Psychological Science in the Public Interest}, 14(1):4--58.
\bibitem[Dwivedi et~al.(2023)]{Dwivedi2023} Dwivedi, Y. K.; et al. 2023. ``So what if ChatGPT wrote it?'' Multidisciplinary perspectives on opportunities, challenges and implications of generative conversational AI. \textit{International Journal of Information Management}, 71:102642. \url{https://doi.org/10.1016/j.ijinfomgt.2023.102642}.
\bibitem[Endsley(1995)]{Endsley1995} Endsley, M. R. 1995. Toward a theory of situation awareness in dynamic systems. \textit{Human Factors}, 37(1):32--64.
\bibitem[Eubanks(2018)]{Eubanks2018} Eubanks, V. 2018. \textit{Automating Inequality: How High-Tech Tools Profile, Police, and Punish the Poor}. New York: St. Martin's Press.
\bibitem[Furze et~al.(2024)]{Furze2024} Furze, L.; Perkins, M.; Roe, J.; and MacVaugh, J. 2024. The AI Assessment Scale (AIAS) in action: A pilot implementation of GenAI-supported assessment. \textit{Australasian Journal of Educational Technology}, 40(4):38--55. \url{https://doi.org/10.14742/ajet.9434}.
\bibitem[Hattie and Timperley(2007)]{HattieTimperley2007} Hattie, J.; and Timperley, H. 2007. The power of feedback. \textit{Review of Educational Research}, 77(1):81--112.
\bibitem[Stephenson and Armstrong(2026)]{HEPI2026} Stephenson, R.; and Armstrong, C. 2026. \textit{Student Generative Artificial Intelligence Survey 2026}. Higher Education Policy Institute, HEPI Report 199. \url{https://www.hepi.ac.uk/reports/student-generative-ai-survey-2026/}.
\bibitem[Hoff and Bashir(2015)]{HoffBashir2015} Hoff, K. A.; and Bashir, M. 2015. Trust in automation: Integrating empirical evidence on factors that influence trust. \textit{Human Factors}, 57(3):407--434.
\bibitem[Jisc(2025)]{Jisc2025} Jisc. 2025. \textit{Student Perceptions of AI 2025}. By Sue Attewell. \url{https://www.jisc.ac.uk/reports/student-perceptions-of-ai-2025}.
\bibitem[Kasneci et~al.(2023)]{Kasneci2023} Kasneci, E.; Sessler, K.; K{\"u}chemann, S.; et al. 2023. ChatGPT for good? On opportunities and challenges of large language models for education. \textit{Learning and Individual Differences}, 103:102274. \url{https://doi.org/10.1016/j.lindif.2023.102274}.
\bibitem[Kestin et~al.(2025)]{Kestin2025} Kestin, G.; Miller, K.; Klales, A.; Milbourne, T.; and Ponti, G. 2025. AI tutoring outperforms in-class active learning: An RCT introducing a novel research-based design in an authentic educational setting. \textit{Scientific Reports}, 15:17458. \url{https://doi.org/10.1038/s41598-025-97652-6}.
\bibitem[Kirschner, Sweller, and Clark(2006)]{Kirschner2006} Kirschner, P. A.; Sweller, J.; and Clark, R. E. 2006. Why minimal guidance during instruction does not work: An analysis of the failure of constructivist, discovery, problem-based, experiential, and inquiry-based teaching. \textit{Educational Psychologist}, 41(2):75--86. \url{https://doi.org/10.1207/s15326985ep4102_1}.
\bibitem[Kulik and Fletcher(2016)]{KulikFletcher2016} Kulik, J. A.; and Fletcher, J. D. 2016. Effectiveness of intelligent tutoring systems: A meta-analytic review. \textit{Review of Educational Research}, 86(1):42--78.
\bibitem[Lee and See(2004)]{LeeSee2004} Lee, J. D.; and See, K. A. 2004. Trust in automation: Designing for appropriate reliance. \textit{Human Factors}, 46(1):50--80.
\bibitem[Lee et~al.(2025)]{Lee2025} Lee, H.-P.; Sarkar, A.; Tankelevitch, L.; Drosos, I.; Rintel, S.; Banks, R.; and Wilson, N. 2025. The impact of generative AI on critical thinking: Self-reported reductions in cognitive effort and confidence effects from a survey of knowledge workers. In \textit{Proceedings of the 2025 CHI Conference on Human Factors in Computing Systems}, Article 1121, 1--22. \url{https://doi.org/10.1145/3706598.3713778}.
\bibitem[Lo(2023)]{Lo2023} Lo, C. K. 2023. What is the impact of ChatGPT on education? A rapid review of the literature. \textit{Education Sciences}, 13(4):410. \url{https://doi.org/10.3390/educsci13040410}.
\bibitem[Long and Magerko(2020)]{LongMagerko2020} Long, D.; and Magerko, B. 2020. What is AI literacy? Competencies and design considerations. In \textit{Proceedings of the 2020 CHI Conference on Human Factors in Computing Systems}, 1--16. \url{https://doi.org/10.1145/3313831.3376727}.
\bibitem[Ma et~al.(2014)]{Ma2014} Ma, W.; Adesope, O. O.; Nesbit, J. C.; and Liu, Q. 2014. Intelligent tutoring systems and learning outcomes: A meta-analysis. \textit{Journal of Educational Psychology}, 106(4):901--918.
\bibitem[Messick(1989)]{Messick1989} Messick, S. 1989. Validity. In Linn, R. L., ed., \textit{Educational Measurement}, 3rd edition, 13--103. New York: Macmillan.
\bibitem[Mittelstadt(2019)]{Mittelstadt2019} Mittelstadt, B. 2019. Principles alone cannot guarantee ethical AI. \textit{Nature Machine Intelligence}, 1(11):501--507.
\bibitem[Ng et~al.(2021)]{Ng2021} Ng, D. T. K.; Leung, J. K. L.; Chu, K. W. S.; and Qiao, M. S. 2021. Conceptualizing AI literacy: An exploratory review. \textit{Computers and Education: Artificial Intelligence}, 2:100041. \url{https://doi.org/10.1016/j.caeai.2021.100041}.
\bibitem[Nicol and Macfarlane-Dick(2006)]{NicolMacfarlaneDick2006} Nicol, D. J.; and Macfarlane-Dick, D. 2006. Formative assessment and self-regulated learning: A model and seven principles of good feedback practice. \textit{Studies in Higher Education}, 31(2):199--218.
\bibitem[Noble(2018)]{Noble2018} Noble, S. U. 2018. \textit{Algorithms of Oppression: How Search Engines Reinforce Racism}. New York: New York University Press.
\bibitem[Nussbaum(2010)]{Nussbaum2010} Nussbaum, M. C. 2010. \textit{Not for Profit: Why Democracy Needs the Humanities}. Princeton, NJ: Princeton University Press.
\bibitem[OECD(2026)]{OECD2026Digital} OECD. 2026. \textit{OECD Digital Education Outlook 2026: Exploring Effective Uses of Generative AI in Education}. Paris: OECD Publishing. \url{https://doi.org/10.1787/062a7394-en}.
\bibitem[Parasuraman and Manzey(2010)]{ParasuramanManzey2010} Parasuraman, R.; and Manzey, D. H. 2010. Complacency and bias in human use of automation: An attentional integration. \textit{Human Factors}, 52(3):381--410. \url{https://doi.org/10.1177/0018720810376055}.
\bibitem[Parasuraman and Riley(1997)]{ParasuramanRiley1997} Parasuraman, R.; and Riley, V. 1997. Humans and automation: Use, misuse, disuse, abuse. \textit{Human Factors}, 39(2):230--253.
\bibitem[Perkins, Roe, and Furze(2025)]{Perkins2025} Perkins, M.; Roe, J.; and Furze, L. 2025. Reimagining the Artificial Intelligence Assessment Scale: A refined framework for educational assessment. \textit{Journal of University Teaching and Learning Practice}, 22(7). \url{https://doi.org/10.53761/rrm4y757}.
\bibitem[Raji et~al.(2020)]{Raji2020} Raji, I. D.; Smart, A.; White, R. N.; Mitchell, M.; Gebru, T.; Hutchinson, B.; Smith-Loud, J.; Theron, D.; and Barnes, P. 2020. Closing the AI accountability gap. In \textit{Proceedings of the 2020 ACM Conference on Fairness, Accountability, and Transparency}, 33--44. \url{https://doi.org/10.1145/3351095.3372873}.
\bibitem[Risko and Gilbert(2016)]{RiskoGilbert2016} Risko, E. F.; and Gilbert, S. J. 2016. Cognitive offloading. \textit{Trends in Cognitive Sciences}, 20(9):676--688.
\bibitem[Roll and Wylie(2016)]{RollWylie2016} Roll, I.; and Wylie, R. 2016. Evolution and revolution in artificial intelligence in education. \textit{International Journal of Artificial Intelligence in Education}, 26:582--599.
\bibitem[Sadler(1989)]{Sadler1989} Sadler, D. R. 1989. Formative assessment and the design of instructional systems. \textit{Instructional Science}, 18:119--144.
\bibitem[Santoni de Sio and van den Hoven(2018)]{Santoni2018} Santoni de Sio, F.; and van den Hoven, J. 2018. Meaningful human control over autonomous systems: A philosophical account. \textit{Frontiers in Robotics and AI}, 5:15. \url{https://doi.org/10.3389/frobt.2018.00015}.
\bibitem[Sarter and Woods(1995)]{SarterWoods1995} Sarter, N. B.; and Woods, D. D. 1995. How in the world did we ever get into that mode? Mode error and awareness in supervisory control. \textit{Human Factors}, 37(1):5--19. \url{https://doi.org/10.1518/001872095779049516}.
\bibitem[Selbst et~al.(2019)]{Selbst2019} Selbst, A. D.; Boyd, D.; Friedler, S. A.; Venkatasubramanian, S.; and Vertesi, J. 2019. Fairness and abstraction in sociotechnical systems. In \textit{Proceedings of the Conference on Fairness, Accountability, and Transparency}, 59--68.
\bibitem[Selwyn(2024)]{Selwyn2024} Selwyn, N. 2024. On the limits of artificial intelligence (AI) in education. \textit{Nordisk tidsskrift for pedagogikk og kritikk}, 10(1):3--14. \url{https://doi.org/10.23865/ntpk.v10.6062}.
\bibitem[Sen(1999)]{Sen1999} Sen, A. 1999. \textit{Development as Freedom}. New York: Knopf.
\bibitem[Sparrow, Liu, and Wegner(2011)]{Sparrow2011} Sparrow, B.; Liu, J.; and Wegner, D. M. 2011. Google effects on memory: Cognitive consequences of having information at our fingertips. \textit{Science}, 333(6043):776--778. \url{https://doi.org/10.1126/science.1207745}.
\bibitem[Sperber et~al.(2010)]{Sperber2010} Sperber, D.; Cl\'{e}ment, F.; Heintz, C.; Mascaro, O.; Mercier, H.; Origgi, G.; and Wilson, D. 2010. Epistemic vigilance. \textit{Mind \& Language}, 25(4):359--393.
\bibitem[Star and Ruhleder(1996)]{StarRuhleder1996} Star, S. L.; and Ruhleder, K. 1996. Steps toward an ecology of infrastructure: Design and access for large information spaces. \textit{Information Systems Research}, 7(1):111--134.
\bibitem[Tai et~al.(2018)]{Tai2018} Tai, J.; Ajjawi, R.; Boud, D.; Dawson, P.; and Panadero, E. 2018. Developing evaluative judgement: Enabling students to make better judgements. \textit{Higher Education}, 76:467--481.
\bibitem[Vallor(2016)]{Vallor2016} Vallor, S. 2016. \textit{Technology and the Virtues: A Philosophical Guide to a Future Worth Wanting}. Oxford: Oxford University Press.
\bibitem[VanLehn(2011)]{VanLehn2011} VanLehn, K. 2011. The relative effectiveness of human tutoring, intelligent tutoring systems, and other tutoring systems. \textit{Educational Psychologist}, 46(4):197--221.
\bibitem[Williamson and Eynon(2020)]{WilliamsonEynon2020} Williamson, B.; and Eynon, R. 2020. Historical threads, missing links, and future directions in AI in education. \textit{Learning, Media and Technology}, 45(3):223--235. \url{https://doi.org/10.1080/17439884.2020.1798995}.
\bibitem[Zawacki-Richter et~al.(2019)]{ZawackiRichter2019} Zawacki-Richter, O.; Mar{\'i}n, V. I.; Bond, M.; and Gouverneur, F. 2019. Systematic review of research on artificial intelligence applications in higher education: Where are the educators? \textit{International Journal of Educational Technology in Higher Education}, 16:39. \url{https://doi.org/10.1186/s41239-019-0171-0}.
\bibitem[Zhang(2026)]{Zhang2026} Zhang, Jianwei. 2026. Intellectual stewardship: Re-adapting human minds for creative knowledge work in the age of AI. \textit{arXiv:2603.18117}. \url{https://arxiv.org/abs/2603.18117}.
\bibitem[University College London(n.d.)]{UCLGenAI} University College London (UCL). n.d. \textit{Three categories of GenAI use in assessment}. Accessed May 10, 2026. \url{https://www.ucl.ac.uk/teaching-learning/generative-ai-hub/three-categories-genai-use-assessment}.
\bibitem[University of Bristol(n.d.)]{BristolAI} University of Bristol. n.d. \textit{Using AI in Assessment}. Accessed May 10, 2026. \url{https://www.bristol.ac.uk/bilt/sharing-practice/guides/guidance-on-ai/using-ai-in-assessment/}.
\bibitem[University of Queensland(n.d.)]{UQAI} University of Queensland. n.d. \textit{UQ's rules for using AI in assessment}. Accessed May 10, 2026. \url{https://itali.uq.edu.au/teaching-guidance/generative-ai-teaching-learning-and-assessment/rules-using-ai-assessment}.
\bibitem[UNSW Sydney(n.d.)]{UNSWAI} UNSW Sydney. n.d. \textit{Guidance on AI in assessment}. Accessed May 10, 2026. \url{https://www.teaching.unsw.edu.au/ai/ai-assessment-guidance}.
\bibitem[University of Auckland(n.d.)]{AucklandAI} University of Auckland. n.d. \textit{Two-Lane Approach to assessment}. Accessed May 10, 2026. \url{https://teachwell.auckland.ac.nz/assessment/two-lane-approach-to-assessment/}.
\bibitem[University of Sydney(2024)]{SydneyAI} University of Sydney. 2024. \textit{University of Sydney's AI assessment policy: protecting integrity and empowering students}. Accessed May 10, 2026. \url{https://www.sydney.edu.au/news-opinion/news/2024/11/27/university-of-sydney-ai-assessment-policy.html}.

\end{thebibliography}
\end{document}